\documentclass{PoS}
\title{Radio-Optical Study of Double-Peaked AGNs. I.~3C~390.3}

\ShortTitle{Inner-jet and optical- emission in 3C390.3 }

\author{\speaker{J.  Le\'on-Tavares}$^1$$^2$\thanks{Member of the CONACyT
    program for PhD studies and visiting student of the International
    Max Planck Research School (IMPRS) for Radio and Infrared Astronomy at the
    Universities of Bonn and Cologne. This work was supported by CONACyT
    research grant 54480 (M\'exico).}, A.P. Lobanov$^2$, V.H.
    Chavushyan$^1$ and T.G. Arshakian$^2$ \\
       \llap{$^1$}Instituto Nacional de Astrof\'{\i}sica, Optica y
 Electr\'onica\\
 AP 51, 72000 Puebla, M\'exico\\ 
        \llap{$^2$} Max-Planck Institut f\"ur Radioastronomie \\
Auf dem H\"ugel 69, 53121 Bonn, Germany \\
      
       E-mail: \email{fleon@inaoep.mx}, \email{alobanov@mpifr-bonn.mpg.de},
       \email{vahram@inaoep.mx}, \email{tigar@mpifr-bonn.mpg.de}}

\abstract{We aim to test the model proposed to explain the correlation between
          the flux density at 15~GHz of a stationary component in the
          parsec-scale jet and the optical continuum emission in the radio
          galaxy 3C~390.3. In the model, the double-peaked emission from
          3C~390.3 is likely to be generated both near the disk and in a
          rotating subrelativistic outflow surrounding the jet, due to
          ionization of the outflow by the beamed continuum emission from the
          jet. This scenario is chosen since broad-emission lines are observed
          to vary following changes in the inner radio jet. For recent epochs
          we have imaged and modelled the radio emission of the inner jet of
          3C~390.3, which was observed with very long baseline interferometry
          at 15~GHz, 22~GHz and 43~GHz, to image the inner part of the
          parsec-scale jet, locate the exact region where the bulk of the
          continuum luminosity is generated and search for the mechanism that
          drives the double-peaked profile emission. We present the
          preliminary results of testing the model using data from 11 years of
          active monitoring of 3C~390.3.}

\FullConference{From planets to dark energy: the modern radio universe\\
		 October 1-5 2007\\
		 University of Manchester, Manchester, UK}

\begin{document}

\section{Inner Jet Structure  in the Double-Peaked AGN 3C~390.3}

Among the variety of types and species in the AGNs ``zoo'', there is a small
fraction of AGNs showing unusual broad and double-peaked profiles of Balmer
and Mg II emission lines, hereafter double-peaked (DP) AGNs. The widths of the
DP profiles range from several thousands to $\sim$~40000~km~s$^{-1}$ . The
first DP emission line profile was discovered in the H${\beta}$ profile of the
broad-line radio galaxy 3C~390.3, whose parsec-scale radio jet has been
monitored at 15 GHz since 1995 with the Very Long Baseline Array (VLBA). The
compact jet can be modelled by circular Gaussian components (see Figure
1). The features \textbf{D} (at \textit{r}=0 mas), \textbf{S1} (at
\textit{r}=0.3 mas) and \textbf{S2} (at \textit{r}=1.5 mas) are stationary
components, whilst the other features are moving. We have measured flux
densities of all the jet components and used back-extrapolation of linear fits
to the component trajectories to calculate the epochs of ejection from the
nucleus (\textbf{D}) and the epochs of passage through the closest stationary
feature (\textbf{S1}). Most of the observations with very long baseline
interferometry (VLBI) have been extracted from the MOJAVE survey [3].

\begin{figure}[ht]
\center
\includegraphics[width=5.5cm,height=5.1cm,angle=0]{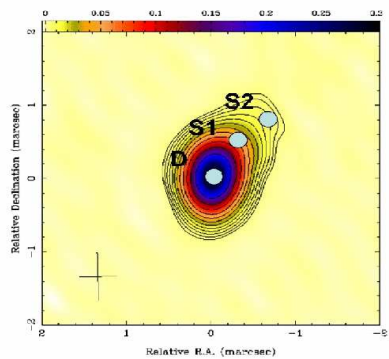}

\caption{\scriptsize{Radio structure of 3C~390.3 Observed in October 2006 with
    the VLBA at 15 GHz. The synthesized beam plotted as a cross in in the
    bottom left corner is 0.56 mas $\times$ 0.44 mas oriented at an angle of
    1.34$^{\circ}$. The peak of the flux density in the image is 304~mJy/beam
    and the rms noise is 0.45 mJy/beam. The contours are drawn at 1,
    $\sqrt{2}$, 2, ... times the lowest contour shown at 0.9 mJy/beam, the
    color scale is in Jy.  The structure observed is quantified by a set of
    two dimensional, circular Gaussian features (\textit{filled circles})
    obtained from fitting the visibility amplitudes and phases. Features
    \textbf{D}, \textbf{S1}, and \textbf{S2} are stationary components. }}
\end{figure}

\section{Results}

Arshakian et al. (2006) found a significant correlation with the flux density
at 15 GHz of the stationary component \textbf{S1} in the parsec-scale jet and
the optical continuum emission in 3C~3903. This strongly implies that the jet
emission from component \textbf{S1} is the main source of the optical
continuum driving the DP emission line variability. We imaged and model fitted
the VLBI data at 15~GHz for recent epochs (2001-2007), finding four new
features (\textbf{C8, C9, C10, C11}). We used measures of the H$\alpha$ broad
line [2] to follow the variations between the jet components and the nuclear
optical emission. We back-extrapolated the trajectories of the features
identified in the jet, the components \textbf{C5-C10} passed through the
location of the stationary feature \textbf{S1} shortly after local $H\alpha$
optical maxima (\textit{green line} in Figure 2). The null hypothesis that
this happens by chance is rejected at a confidence level of 99.99~\% . The
H$\alpha$ light curve shows a correlation with the flux density of the
stationary feature \textbf{S1} at a high confidence level, whilst there is no
such correlation for any other component. In Figure 2 is shown the apparent
anticorrelation between the flux variability of the component \textbf{D} at
the base of the jet and the stationary component \textbf{S1} becoming dramatic
for the most recent epochs.

\begin{figure}[ht]
\center \includegraphics[width=6cm,height=8.cm,angle=90,clip]{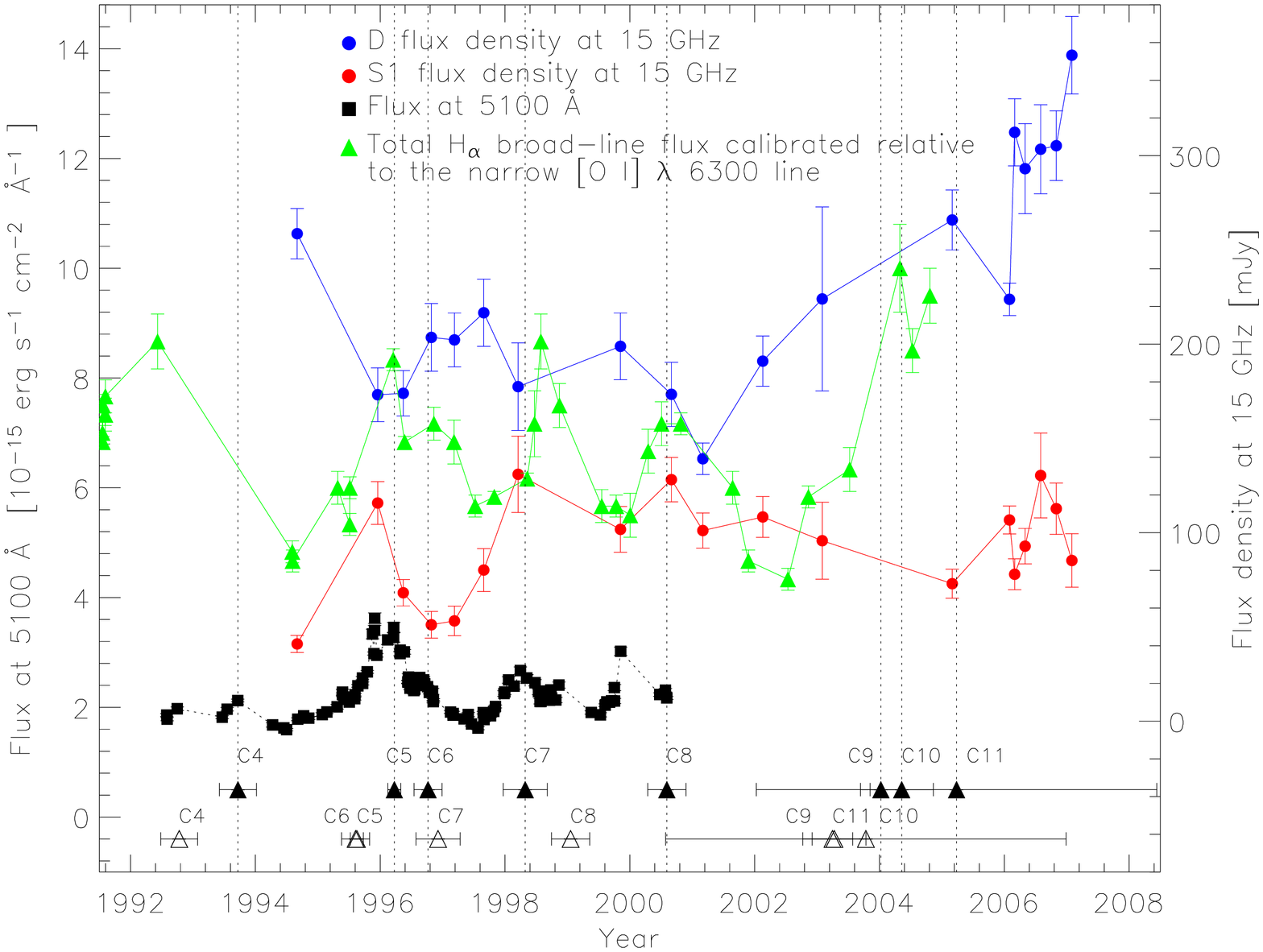} 
\includegraphics[width=7cm,height=6cm,angle=0,clip]{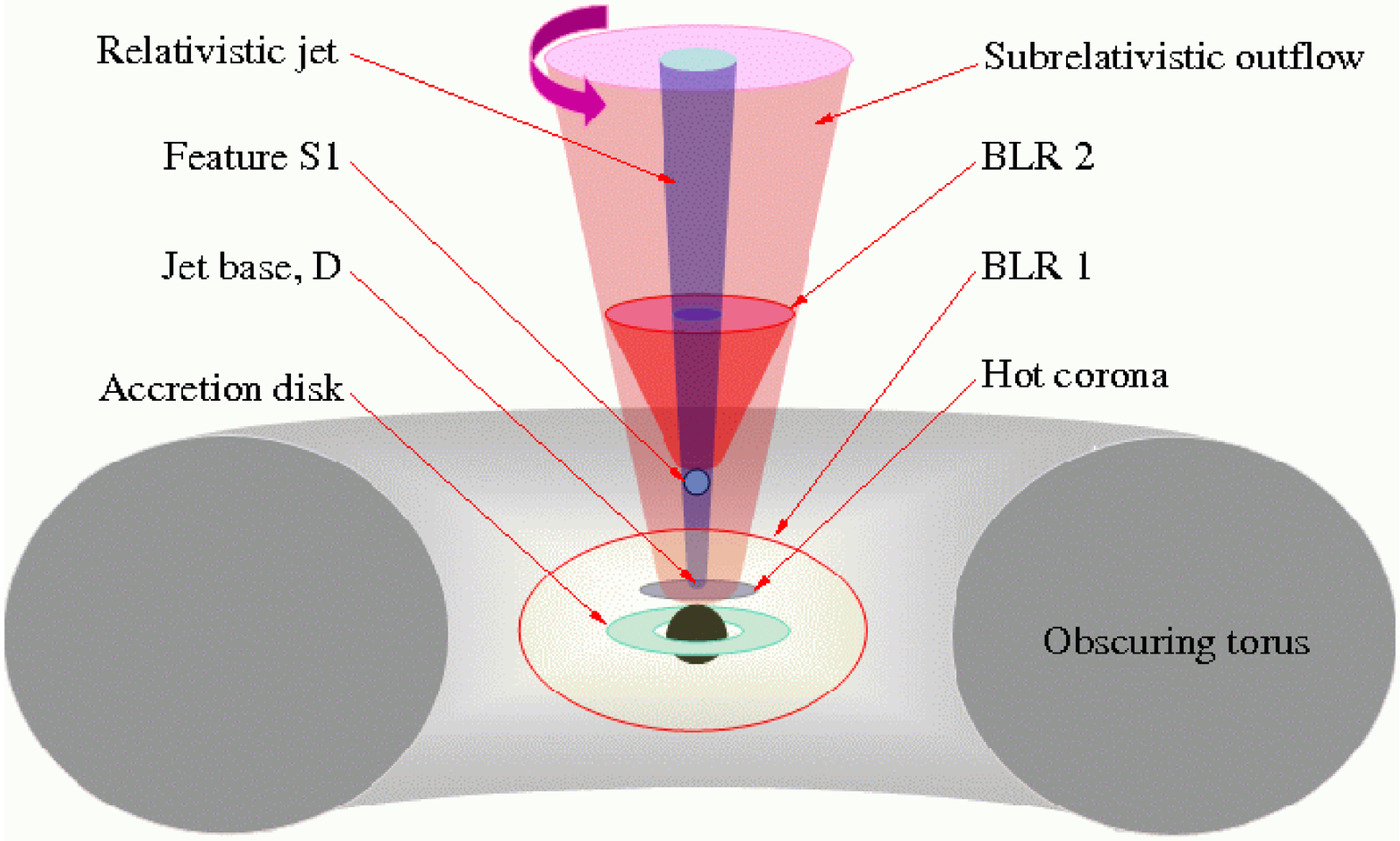}
\caption{\scriptsize{\textbf{Left panel:} Flux variations of the inner jet
    stationary components and the optical emission. \textit{Black-squares}:
    Optical continua flux at 5100 \AA [4]. \textit{Green-triangles}: Total
    H$\alpha$ broad-line flux calibrated relative to the narrow
    [O I] $\lambda$6300 line [2] (the flux scales in the plot do not apply to
    this measures). \textit{Filled-circles}:  Flux density of the inner-jet
    features \textbf{D} (\textit{blue}) and \textbf{S1} (\textit{red}). The ejection
    epochs (\textit{open-triangles}), and the epochs of passages through the
    stationary feature (\textit{filled-triangles}) \textbf{S1}. \textbf{Right~
    panel:} A sketch of the nuclear region in 3C~390.3 ( the drawing is not to
    scale and shows only the approaching jet). The broad-line emission is
    likely to be generated both near the disk (BLR1, ionized by the emission
    from a hot corona or the accretion disk) and in a rotating subrelativistic
    outflow surrounding the jet (BLR2, ionized by the emission from the
    relativistic plasma in the jet). BLR2 is evident in the broad-line
    emission near the maxima in the optical light curve, when the jet emission
    dominates the optical continuum. BLR1 may be manifested in the broad-line
    emission around the epochs of minima in the optical flux, when the jet
    contribution to the ionizing continuum is small.}}
\end{figure}

\section{Discussion and Future work}

We found a correlation between the \textbf{S1} jet component and the H$\alpha$
optical emission using 11 years of VLBI and H$\alpha$ broad-emission line
monitoring, supporting the idea that the optical nuclear emission in 3C~390.3
has a non-thermal origin.  Figure 2 shows the suggested scheme that explains
the basic properties of the optical emission (continuum and broad H$\alpha$)
driven by the inner jet.  The multi-frequency analysis and the
spectrophotometric monitoring for 3C 390.3 still in progress.

\end{document}